\documentclass[aps,
twocolumn,
superscriptaddress]{revtex4}
\usepackage{amsfonts}
\usepackage{amsmath}
\usepackage{amssymb}
\usepackage{graphicx}
\usepackage[dvipsnames]{xcolor}
\def\e{$\varepsilon$}

\begin{document}

\title{Magnetic fluctuations and superconducting pairing in $\varepsilon$-iron}

\author{A. S. Belozerov}
\affiliation{M. N. Miheev Institute of Metal Physics, Russian Academy of Sciences, 620108 Yekaterinburg, Russia}
\affiliation{Ural Federal University, 620002 Yekaterinburg, Russia}

\author{A. A. Katanin}
\affiliation{Moscow Institute of Physics and Technology, 141701 Dolgoprudny, Russia}
\affiliation{M. N. Miheev Institute of Metal Physics, Russian Academy of Sciences, 620108 Yekaterinburg, Russia}

\author{V. Yu. Irkhin}
\affiliation{M. N. Miheev Institute of Metal Physics, Russian Academy of Sciences, 620108 Yekaterinburg, Russia}
\affiliation{Ural Federal University, 620002 Yekaterinburg, Russia}

\author{V. I. Anisimov}
\affiliation{M. N. Miheev Institute of Metal Physics, Russian Academy of Sciences, 620108 Yekaterinburg, Russia}
\affiliation{Ural Federal University, 620002 Yekaterinburg, Russia}

\begin{abstract}
We study Coulomb correlation effects and their role in superconductivity of $\varepsilon$-iron under pressure from 12 to 33 GPa by using a combination of density functional and dynamical mean-field theory. Our results indicate a persistence of the Fermi-liquid behavior below the temperature $\sim$1000~K. The Coulomb correlations are found to substantially renormalize the density of states, reducing the distance from the peak to the Fermi level to 0.4 eV compared to 0.75 eV obtained in DFT calculations. We find significant antiferromagnetic correlations, which are accompanied by the formation of short-lived local magnetic moments. We use the obtained results as a starting point for construction of the multi-band Bethe-Salpeter equation, which eigenvalues indicate that antiferromagnetic spin fluctuations may result in the superconducting pairing in $\varepsilon$-Fe. Moreover, the tendency to superconducting instability becomes weaker with the increase of pressure, which may explain the disappearance of superconductivity at $\sim$30 GPa.

\end{abstract}
\maketitle

\section{Introduction}


Since the discovery of the hexagonal close-packed (hcp) iron ($\varepsilon$ phase) in 1964~\cite{Takahashi1964}, its puzzling properties 
have attracted significant interest.
This phase appears at pressures above 12~GPa and, as shown by static compression experiments,
is stable up to pressures and temperatures corresponding to Earth's core conditions~\cite{Tateno2010}.
It makes the investigation of $\varepsilon$-Fe is particularly relevant for geophysics and Earth's magnetism~\cite{Pourovskii2013,Hausoel}.

One of the actively debated
characteristics of \e-Fe is its magnetic behavior.
The M\"ossbauer effect measurements found 
no hyperfine magnetic field at temperatures down to 30~mK~\cite{mossbauer_0,mossbauer_1,mossbauer_2,mossbauer_3},
implying that hcp iron is non-magnetic. 
The static magnetic order was not also detected by x-ray magnetic circular dichroism (XMCD)~\cite{Mathon2004} and
neutron powder diffraction~\cite{XES3}.
The anomalous mode splitting in the Raman spectra~\cite{Merkel2000}, interpreted initially as a hint of the presence of magnetic order, was later shown to be inconsistent with it~\cite{Goncharov2003}.
At the same time, the x-ray emission spectroscopy (XES) consistently detected remnant magnetism, decreasing with pressure up to $\sim$35~GPa~\cite{XES1,XES2,XES3}.
%
%

The presence of 
magnetic correlations in \mbox{\e-Fe} is supported by density functional theory (DFT) studies.
Most of early calculations predicted the antiferromagnetic type-II (AFM-II) ground state, which is stable up to $\sim$50~GPa~\cite{dft_1,dft_2,dft_3,dft_6}, or the spin-spiral ground state~\cite{Thakor2003,dft_4}.
It is worth noting that 
AFM-II structure leads to a better agreement of structural and elastic properties with experiments, as compared to non-magnetic calculations~\cite{dft_1,dft_2}.
%
%
%
%
Nevertheless, recent DFT study of Lebert \textit{et al.} revealed three intensity-modulated
phases with a lower energy than AFM-II state~\cite{XES3}.
In this study, the ground state at pressures 15$-$35~GPa was found to be formed by alternating AFM and non-magnetic bilayers.

At the same time, the \textit{paramagnetic} state of \e-Fe was studied by a combination of DFT and dynamical mean-field theory (DFT+DMFT)~\cite{DFT+DMFT}.
These studies showed the importance of Coulomb correlations for description of the 
electronic~\cite{Glazyrin2013,Pourovskii2014} and structural~\cite{Pourovskii2014} properties at moderate pressures, as well as phase stability~\cite{Pourovskii2013}, transport~\cite{Pourovskii2017,Xu2018,Zhang2015} and magnetic~\cite{Pourovskii2013,Vekilova2015} properties at Earth's core conditions.
In addition, a jump of resistivity at the ${\alpha\!\rightarrow\!\varepsilon}$ transition was qualitatively captured by Pourovskii \textit{et al.}~\cite{Pourovskii2014},
while quantitative difference was attributed to non-local correlation effects (for a review, see Ref.~\cite{Pourovskii2019}).

The properties of \e-Fe have become even more intriguing with the discovery of superconductivity in the pressure range from 15 to 30~GPa, with the largest critical temperature ${T_\textrm{c}=2}$~K
at about 21~GPa~\cite{Shimizu2000}.
%
%
%
The superconducting properties are also unusual. In particular, the superconductivity seems to occur in paramagnetic phase; the large slope 
of the critical field curve at $T_\textrm{c}$
demonstrates that heavy particles 
are involved in the 
pairing (see Ref.~\cite{Holmes2004} and review in Ref.~\cite{Flouquet}).
%
%
%
Moreover, the temperature dependence of resistivity shows a $T^{5/3}$ power-law behavior below 30~K over the entire superconducting pressure region~\cite{Jaccard2002,Holmes2004,Jaccard2005,Pedrazzini2006,Sengupta2010,Yadav2013}. This non-Fermi-liquid behavior (in contrast with the $T^{2}$ dependence for Fermi liquid) was attributed to ferromagnetic (FM) fluctuations according to Moriya's theory of weak itinerant magnetism~\cite{Moriya}.
%
%
%
These experimental results point to
an unconventional superconductivity; 
the presence of FM fluctuations may be a signature of triplet
pairing symmetry.
However, it is at odds with the DFT calculations, predicting a proximity to antiferromagnetism, which would likely yield singlet superconducting state.
%
%
%

A narrow pressure range of superconductivity also suggests that it has a spin-fluctuation origin. 
This was confirmed by calculations of the electron-phonon coupling within DFT, which showed that the phonon mechanism can explain the appearance of superconductivity in \e-Fe, but not its rapid disappearance 
with pressure~\cite{Mazin,Bose2003}.
%
%
In particular, the calculated range of superconductivity
was found to extend
well above 100~GPa.
%
%
%
To consider the magnetic pairing mechanism, Jarlborg estimated the coupling constant within DFT and concluded that superconductivity mediated by spin fluctuations is more likely than that owing to electron-phonon interaction~\cite{Jarlborg}.





%

Above discussed observations stress the importance of studying correlation effects in $\varepsilon$-iron. In this paper, we study the electronic and magnetic properties of paramagnetic \e-Fe within superconducting pressure range by employing DFT+DMFT approach. 
%
%
%
We find considerable local correlations, leading to formation of short-lived local magnetic moments, as well as pronounced incommensurate spin fluctuations. 
To clarify whether these fluctuations can mediate the electron pairing, we derive and solve the multi-band Bethe-Salpeter equation (BSE) in the particle-particle channel.
%
The obtained results indicate that the spin fluctuations in \e-Fe may lead to superconducting instability in the experimental pressure range.

%

The plan of the paper is the following. In Sec. II, we list the computational details. In Sec. III, we present the results on electronic, magnetic and superconducting properties, and finally, in Sect. IV, the conclusions are given.
In Appendix, we derive the gap equation and pairing interaction.

\section{Computational details}
\label{sec:computational_details}

We have performed DFT calculations of $\varepsilon$-iron using the full-potential linearized augmented-plane wave method implemented in the ELK code supplemented by the Wannier function projection procedure. 
The Perdew-Burke-Ernzerhof form of generalized gradient approximation (GGA) was considered.
The calculations were carried out with the experimental lattice constants at the corresponding pressures~\cite{Dewaele2006}.
The convergence threshold for total energy was set to $10^{-6}$~Ry.
The integration in the reciprocal space was performed using 22$\times$22$\times$12 $\textbf{k}$-point mesh in all calculations except those of the Fermi surface and momentum-dependent susceptibility, where 42$\times$42$\times$22 mesh was employed.
From converged DFT results we have constructed effective Hamiltonians in the basis of Wannier functions, which were built as a projection of the original Kohn-Sham states to site-centered localized functions as described in Ref.~\onlinecite{Korotin08}, considering $3d$, $4s$ and $4p$ states.

In DMFT calculations we use for $d$-states the Hubbard parameter 
${U \equiv F^0=4.3}$~eV
and Hund's rule coupling
${I=1.0}$~eV,
which were used in previous study of \e-Fe
and resulted in accurate descriptions of its structural properties at the same pressures as in current work~\cite{Pourovskii2014}.
We note that these values are in overall agreement with estimates for $\alpha$-iron~\cite{Belozerov2014}. 
Our calculations have been performed using the AMULET code~\cite{amulet}.
%
The double-counting correction has been taken in the 
around mean-field form.
We also verified that the fully localized double-counting correction leads to similar results, although providing a slightly larger ($\sim$0.1) filling of \textit{d} states.
%
%
The impurity problem has been solved by the hybridization expansion continuous-time quantum Monte Carlo method~\cite{CT-QMC} with the density-density form of Coulomb interaction.
We consider the redistribution of charge density, caused by the self-energy, only within DMFT self-consistency loop. We checked this approximation at several pressures and temperatures by performing DFT+DMFT calculations both with and without full charge self-consistency, which led to similar results.


We note that the DMFT takes into account local electronic correlations, neglecting the momentum-dependence of self-energy~\cite{dmft}. Although the nonlocal corrections to DMFT can be obtained using cluster~\cite{clusters} or diagrammatic~\cite{OurRevModPhys} methods, these approaches are too computationally expensive to be applied to
real multiorbital compounds at the moment.

\section{Results and discussion}
\subsection{Electronic properties\label{Sect:elProp}}

\begin{figure}[b]
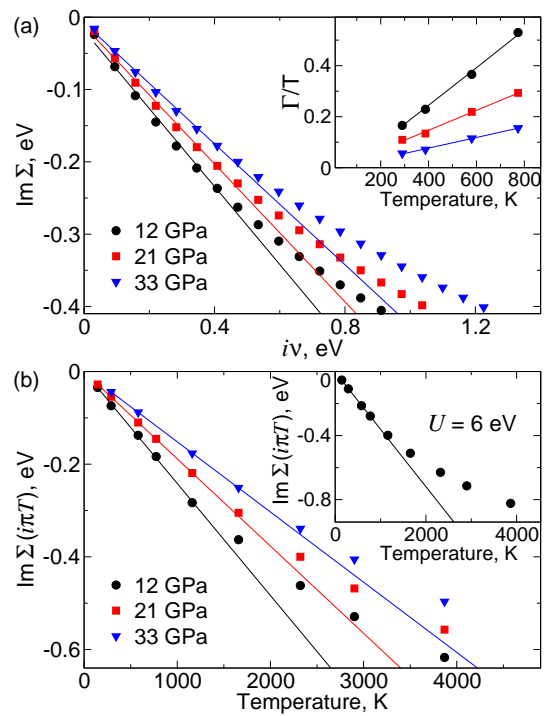

\includegraphics[clip=true,width=0.393\textwidth]{fig1a_Sigma2_beta100_np697_v2.eps}
\includegraphics[clip=true,width=0.393\textwidth]{fig1b_Sigma_at_first_Matsubara3.eps}
\caption{
\label{sigma_and_damping}
(a) Imaginary part of self-energy for degenerate $xy$ and ${x^2{-}y^2}$ states as a function of imaginary frequency $i\nu$ obtained by DFT+DMFT method at ${\beta=100}$~eV$^{-1}$. Inset: temperature dependence of the ratio of quasiparticle damping $\Gamma$ to temperature $T$. 
(b) Imaginary part of the above mentioned self-energy at the first Matsubara frequency as a function of temperature.
Inset: the same for ${U=6}$~eV.
The straight lines depict the fit to linear dependence.}
\end{figure}

Within the DFT+DMFT we find the filling of various $d$-orbitals in the range ${n=1.30\div1.36}$,
which is close to the values in $\alpha$- and $\gamma$-iron \cite{alpha_iron2010,OurGamma,alpha_iron2017}, but the peak of the density of states is sufficiently far from the Fermi level (see below).
We first analyze possible deviations from the Fermi-liquid (FL) regime by using several criteria employed earlier at Earth's core conditions~\cite{Pourovskii2013,Pourovskii2017}.
%
For brevity, we present the results only for degenerate $xy$ and ${x^2{-}y^2}$ states, while similar results were obtained for $z^2$ and degenerate $xz$, $yz$ states.

In a FL state, the imaginary part of self-energy depends linearly 
on imaginary frequency $i\nu_n$ at small $|\nu_n|$ as
${\textrm{Im}\, \Sigma(i\nu_n)\approx -\Gamma -({Z^{-1} {-}1})\nu_n}$, where
$\Gamma$ is the quasiparticle damping (i.e., inverse quasiparticle lifetime),
$Z$ is the 
quasiparticle residue.
%
In Fig.~\ref{sigma_and_damping}(a) we show the obtained imaginary part of $\Sigma(i\nu_n)$, which scales almost linearly for all considered pressures.
To obtain a more quantitative estimate, we compute the quasiparticle damping $\Gamma$, which in a FL depends quadratically on temperature, implying $\Gamma/T\propto T$.
%
%
To this end, 
we perform the analytical continuation of self-energies by using the Pad\'e approximants~\cite{Pade}.
The obtained temperature dependence of $\Gamma/T$, shown in the inset of Fig.~\ref{sigma_and_damping}(a), also 
corresponds to the coherent FL regime.

\begin{figure}[t]
\includegraphics[clip=true,width=0.385\textwidth]{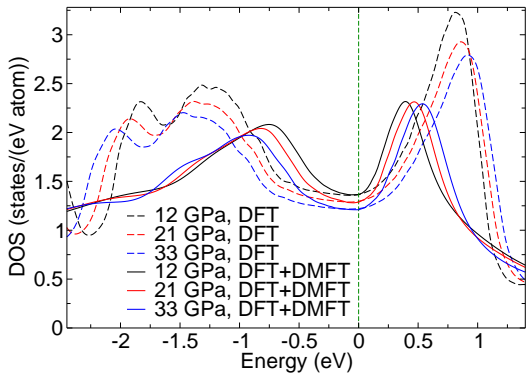}
\caption{
\label{fig:dos}
Density of states of $\varepsilon$-Fe obtained within DFT and DFT+DMFT at 
${T=290}$~K
for different pressures. The Fermi level is at zero energy.}
\end{figure}

%
To obtain an accurate estimate of the temperature $T^*$, 
below which the crossover to the FL regime occurs, avoiding 
use of Pad\'e approximants,
we consider the ``first Matsubara frequency" rule~\cite{Chubukov2012}, according to which in the local approximation ${\textrm{Im}\, \Sigma}$ at first Matsubara frequency ${\nu_1=\pi T}$
does not contain any terms beyond $\mathcal{O}(T)$ in the FL regime. 
%
%
As shown in Fig.~\ref{sigma_and_damping}(b), ${\textrm{Im}\, \Sigma (i\pi T)}$ is proportional to temperature below 
$T^*$
which is $\sim$1500~K at 12 GPa. 
%
A gradual increase of $T^*$ with pressure reflects the reduction of the Coulomb correlation strength due to larger bandwidth. 
These results suggest a possibility of suppression of coherence temperature $T^*$
for 
larger Hubbard~$U$.
However, our calculations with ${U=6}$~eV lead to only a weak change 
of $T^*$, which is $\sim$1000~K at pressure of 12~GPa (see inset in Fig.~\ref{sigma_and_damping}(b)).
%
The value of $T^{*}$ can be also affected by 
using the full rotationally-invariant Coulomb interaction,
while the density-density approximation was shown to underestimate the magnitude of scattering ${|\textrm{Im}\, \Sigma (i\nu_n)|}$ in \e-Fe~\cite{Pourovskii2019}.
%
%
The FL behavior of \e-Fe was also predicted by Pourovskii \textit{et al.} at Earth's core conditions~\cite{Pourovskii2013,Pourovskii2017}, even at the relatively high temperature of $\sim$6000~K, while 
a contrary conclusion was obtained by Zhang~\textit{et al.}~\cite{Zhang2015}.

%
%

The Coulomb correlations in \e-Fe result in substantial mass enhancement ${m^*/m=Z^{-1}}$, which, averaged over all $d$ orbitals, is equal to 1.8 at 12 GPa and gradually decreases to 1.55 at 33 GPa in agreement with earlier studies~\cite{Pourovskii2014}.
The enhancement of the quasiparticle mass is accompanied by the renormalization of the density of states (DOS) near the Fermi level as displayed in Fig.~\ref{fig:dos}. As a result, the distance from the peak lying above the Fermi level is decreased to 0.4~eV compared to 0.75~eV obtained in DFT calculations at 12 GPa. The DOS at the Fermi level decreases with pressure, but not rapidly enough to explain the disappearance of superconductivity, as was also noted previously~\cite{Mazin}.



\subsection{Magnetic properties}

\begin{figure}[b]
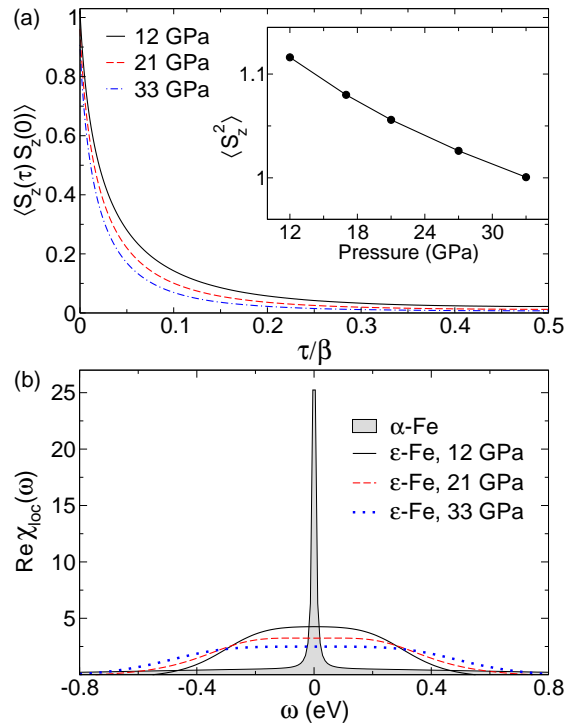

\includegraphics[clip=true,width=0.41\textwidth]{fig3a_spin_corr5_with_Sz2.eps}
\includegraphics[clip=true,trim=-0.05cm 0cm 0cm 0cm,width=0.41\textwidth]{fig3b_spin_corr_w_with_alpha_iron.eps}
\caption{
\label{fig:correlators}
Local spin-spin correlation functions in the imaginary-time (a) and real energy (b) domains calculated by DFT+DMFT method at ${\beta=40}$~eV$^{-1}$. In the bottom panel, the obtained correlation functions are compared with that for $\alpha$-Fe~\cite{footnote1}. 
Inset: instantaneous average $\langle S_z^2 \rangle$ for \e-Fe as a function of pressure.}
\end{figure}

In Fig.~\ref{fig:correlators}(a) we present the local spin-spin correlation functions ${\langle S_z(\tau) S_z(0)\rangle}$ at ${T=290}$~K, which 
have a significant instantaneous average ${\langle S_z^2 \rangle\simeq1}$, and decay rapidly with imaginary time $\tau$.
This $\tau$-dependence
corresponds to 
fast quantum fluctuations of instantaneous magnetic moments on lattice sites.
%
To estimate the degree of spin localization,
%
we Fourier transform the spin correlator to imaginary bosonic frequencies,  $\chi_\textrm{loc}(i\omega_n)=\int_0^\beta \exp(i\omega_n\tau){\langle S_z(\tau) S_z(0)\rangle}$, and then analytically continue it to real frequency $\omega$. The real part of the obtained $\chi_\textrm{loc}(\omega)$ is shown in Fig.~\ref{fig:correlators}(b) in comparison with that of $\alpha$-iron~\cite{footnote1} (see also Refs.~\cite{alpha_iron2010,OurGamma,alpha_iron2017}), being a system with well-defined local moments.
The half width of the peak in Re$\chi_\textrm{loc}(\omega)$
at half of its height yields approximately inverse lifetime of local magnetic moments~\cite{OurGamma}. For $\varepsilon$-iron we therefore obtain the inverse lifetime in energy units ranging from $0.3$~eV
at $p=12$~GPa to $0.5$~eV at $p=33$~GPa, which corresponds to the lifetime of local moments ranging from $14$ to $8$~fs, respectively. This is comparable to the lifetime of local moments, discussed previously in pnictides \cite{Pnictides}.
Thus the obtained results show a presence of short-lived local magnetic moments, which lifetime decreases with increasing pressure in agreement with XES measurements~\cite{XES1,XES2,XES3}.

\begin{figure}[t]
\includegraphics[clip=true,width=0.4\textwidth]{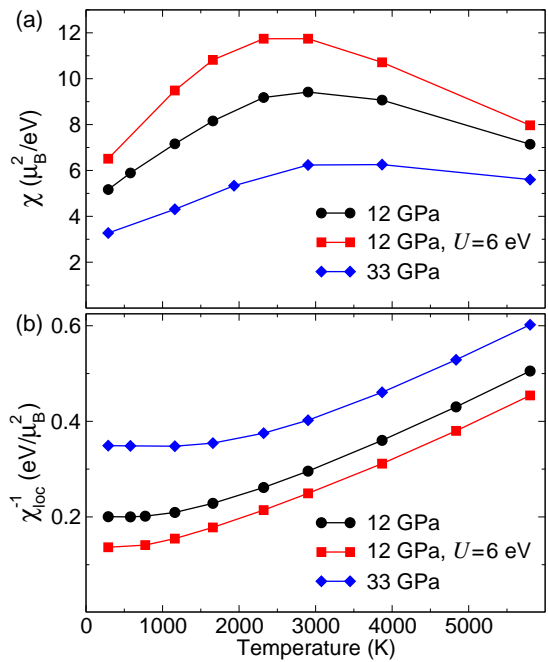}
\caption{
Temperature dependence of uniform magnetic susceptibility (a)
and inverse local susceptibility (b) for \e-Fe obtained within DFT+DMFT method. 
\label{fig:susc}
}
\end{figure}

To get further insight into the formation of local
moments, we calculate the uniform magnetic susceptibility
as a response to a small external magnetic field introduced in the DMFT part (see Fig.~\ref{fig:susc}(a)).
In particular, we have used the magnetic field corresponding to splitting of the single-electron energies by 0.02~eV, which was checked to provide a linear response.
The calculated uniform susceptibility, shown in Fig.~\ref{fig:susc}(a), increases almost linearly with temperature at low temperatures and decrease above a certain temperature $\sim3000$ K, determined by the distance from the peak of the density of states to the Fermi level, similarly to other systems with a peak of the density of states shifted off the Fermi level \cite{OurGamma,LaFeAsO}. 
Increasing value of $U$ leads to increase of uniform susceptibility.

The static 
local spin susceptibility ${\chi}_{\rm loc}=  4\mu_\textrm{B}^2{\chi}_{\rm loc}(0)$ 
shows a Pauli-like
behavior below the temperature $T^*$ of a crossover to the Fermi liquid regime
and a Curie-Weiss behavior ${\chi_{\rm loc} \propto (T-T_{\rm loc}})^{-1}$ above $T^*$ due to presence of the unscreened local moments (see Fig.~\ref{fig:susc}(b)).
The ``Weiss" temperature $T_{\rm loc}$ is negative and determines the Kondo temperature ${T_\textrm{K}\sim -T_{\rm loc}}$ below which the local moments are screened by conduction electrons, similarly to the single-impurity Kondo model \cite{Wilson,Melnikov,Tsvelik}, for which ${T_\textrm{K}\approx -T_{\rm loc}/\sqrt{2}}$. We find ${T_{\rm loc}\sim -T^*}$ and therefore 
rather large Kondo temperatures 
${T_\textrm{K}\sim T^*}$. Similarly to $T^*$, the Kondo temperature is somewhat suppressed at ${U=6}$~eV.

\begin{figure}[t]
\includegraphics[clip=true,width=0.415\textwidth]{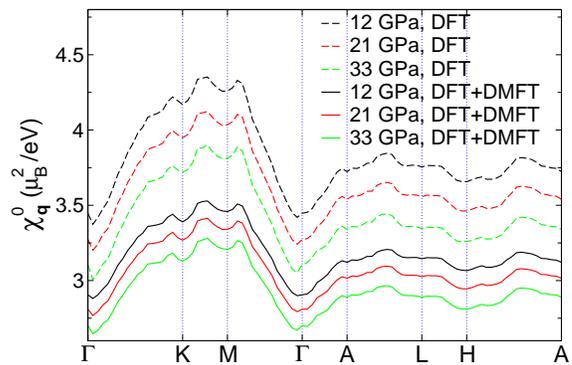}
\caption{\label{fig:susc_irr}
Momentum-dependence of the particle-hole bubble
obtained within DFT and DFT+DMFT at ${\beta=100}$~eV$^{-1}$.}
\end{figure}

Next we calculate the particle-hole bubble
\vspace{-0.1cm}
\begin{equation}
\chi_\textbf{q}^{0} = -(2\mu_B^2/\beta) \sum_{\textbf{k},\nu_n} \textrm{Tr}[G_\textbf{k} (i\nu_n) G_\textbf{k+q}(i\nu_n)],
\end{equation}
where 
$G_\textbf{k}(i\nu_n)$ is the one-particle Green function, which is a matrix in the $d$-orbital space, obtained using the Wannier-projected Hamiltonian and $\nu_n$ are the Matsubara frequencies.
In Fig.~\ref{fig:susc_irr} we show the momentum dependence of $\chi_\textbf{q}^{0}$ calculated using non-interacting (DFT) and interacting (DFT+DMFT) propagators.
In both cases, $\chi_\textbf{q}^{0}$
has a similar shape with a maximum at incommensurate wave vector, lying near K and M points  of the Brillouin zone. These preferable wave vectors are in agreement with previous DFT analysis \cite{dft_4,dft_6,Thakor2003}. The particle-hole bubble increases with decreasing pressure due to decrease of the bandwidth. Therefore, both uniform and non-uniform spin fluctuations become stronger with decreasing pressure.

\subsection{Superconducting pairing}

\begin{figure*}[t]
\includegraphics[clip=true,width=0.242\textwidth]{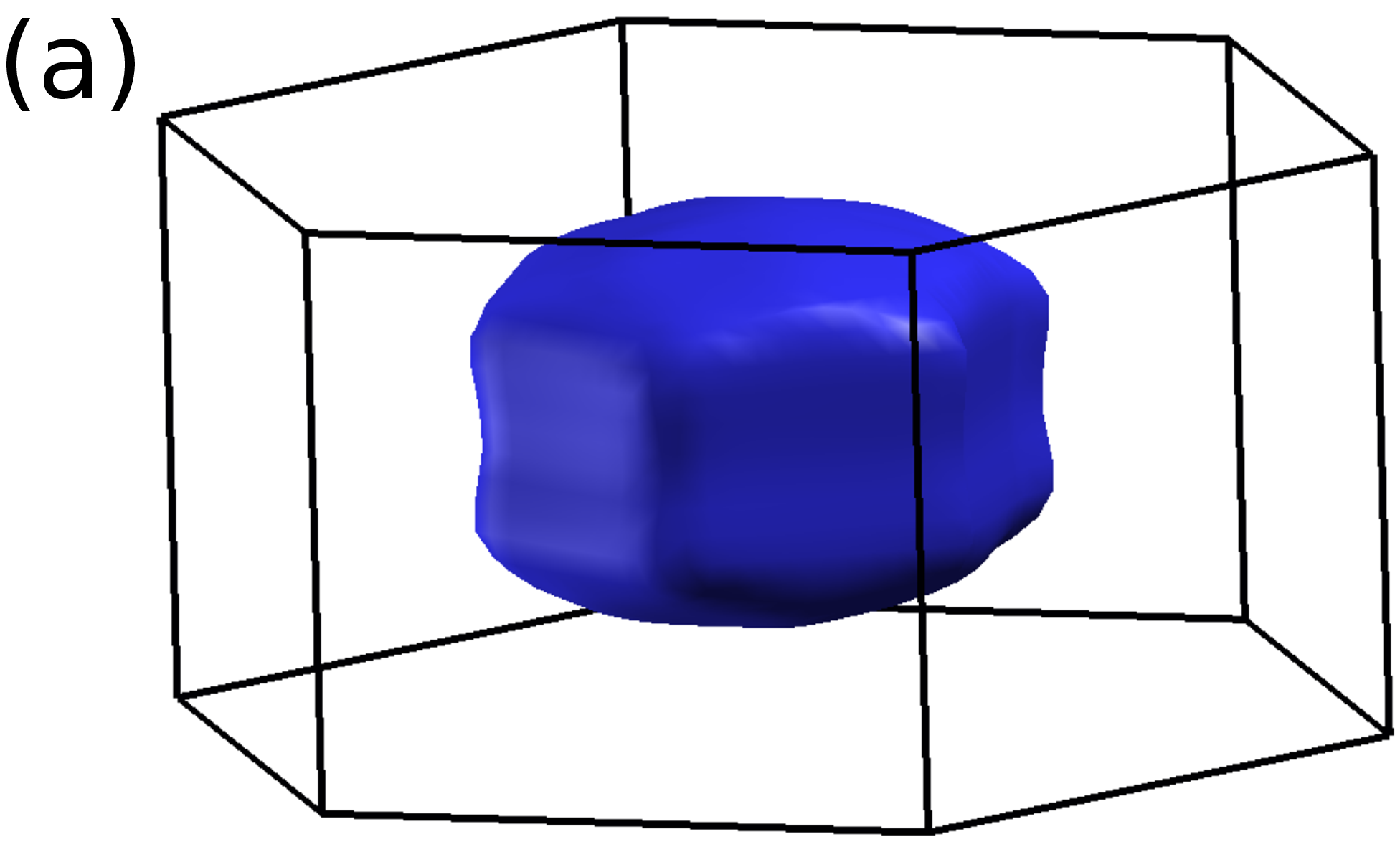}
\hspace{0.4cm}
\includegraphics[clip=true,width=0.242\textwidth]{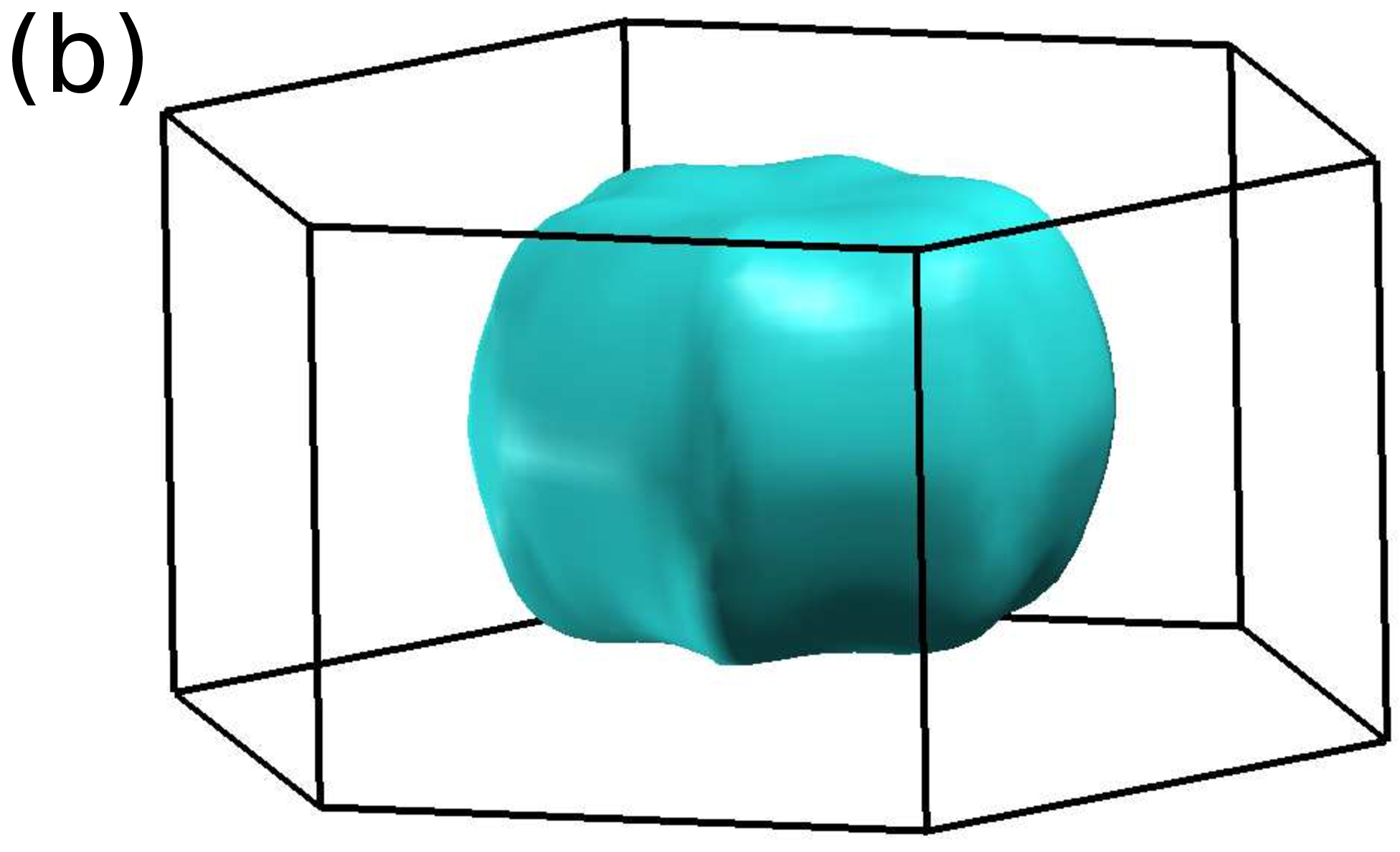}
\hspace{0.4cm}
\includegraphics[clip=true,width=0.242\textwidth]{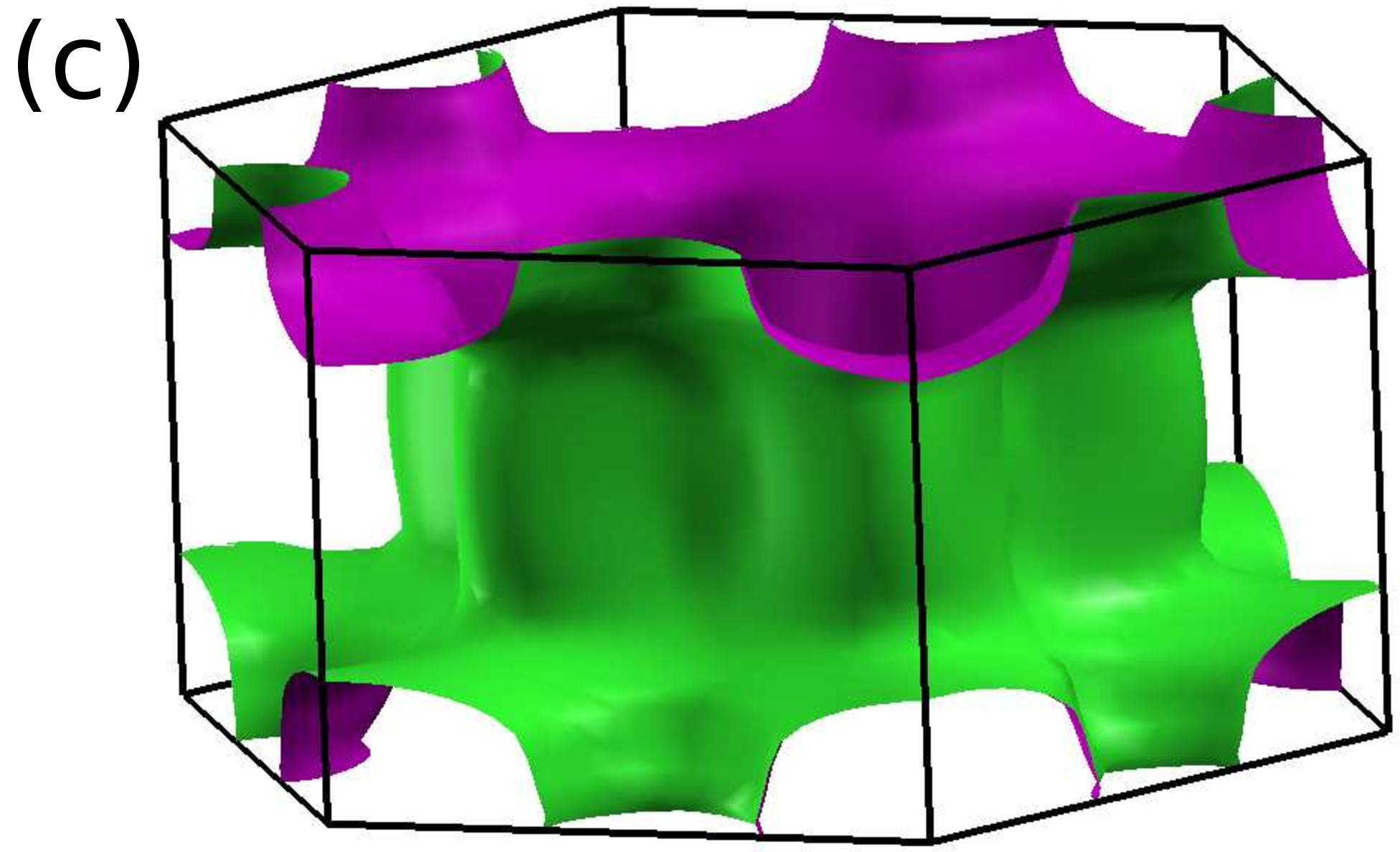}\\
\hspace{0.45cm}
\includegraphics[clip=true,width=0.24\textwidth]{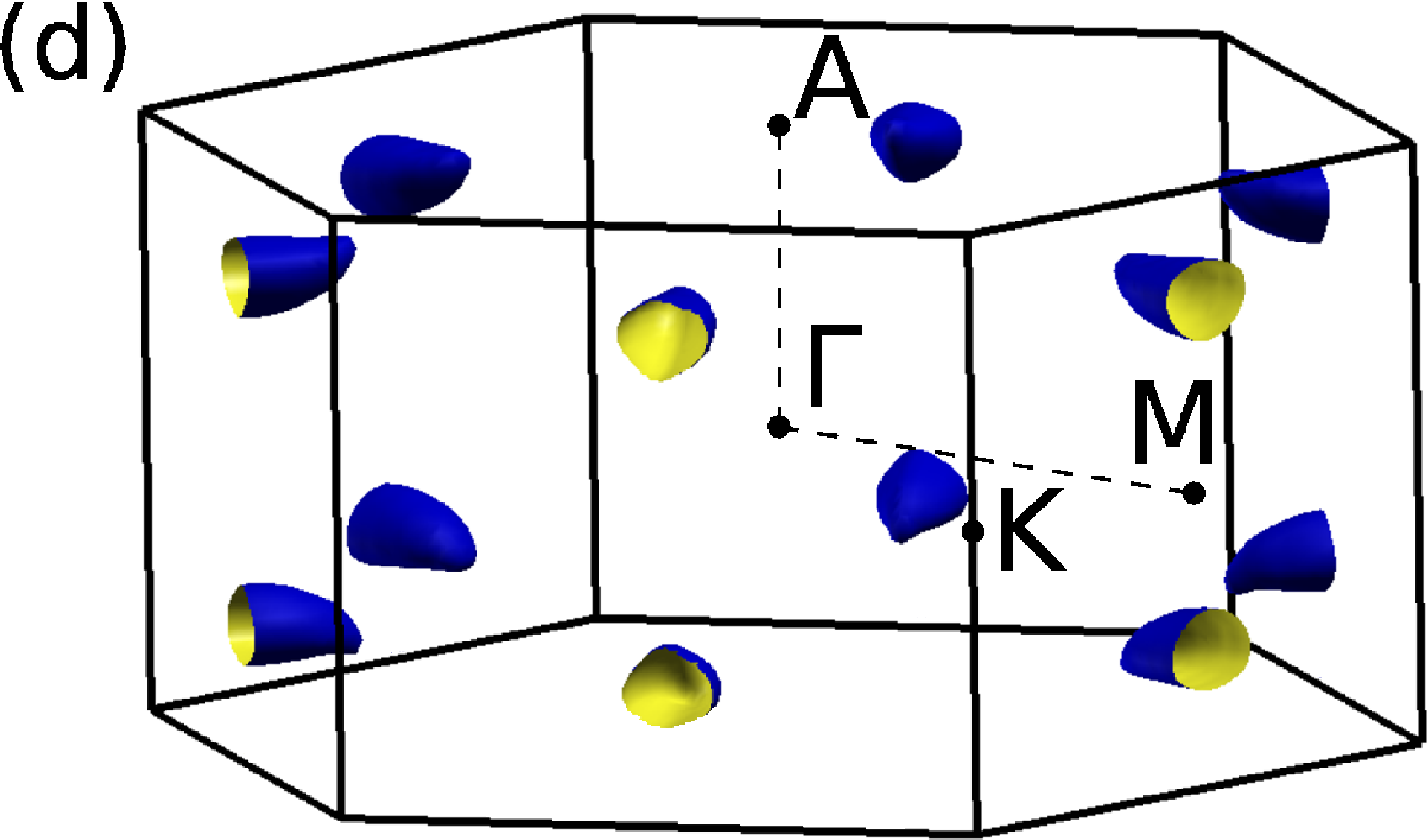}
\hspace{0.5cm}
\includegraphics[clip=true,width=0.24\textwidth]{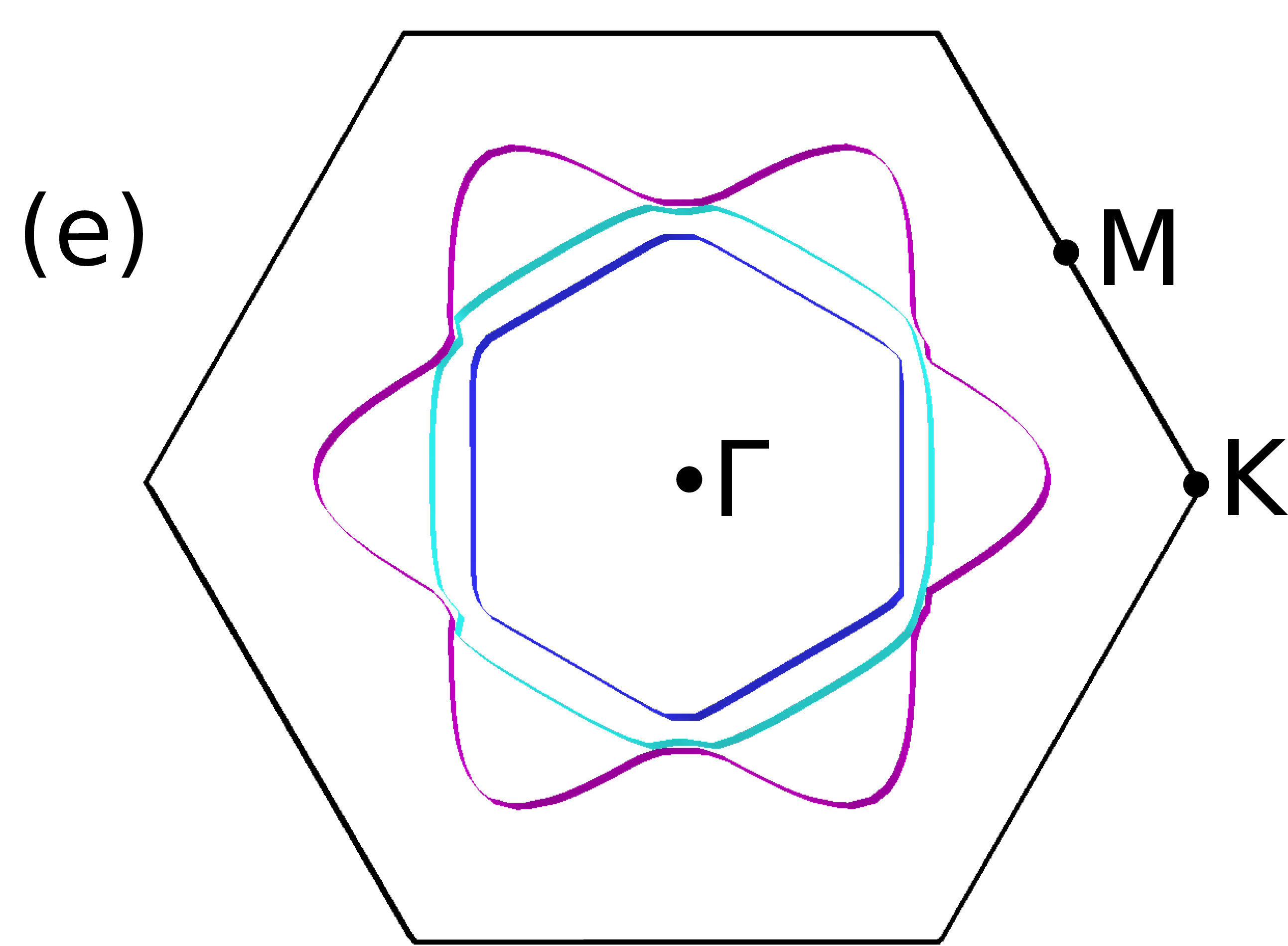}
\hspace{0.25cm}
\includegraphics[clip=true,width=0.271\textwidth]{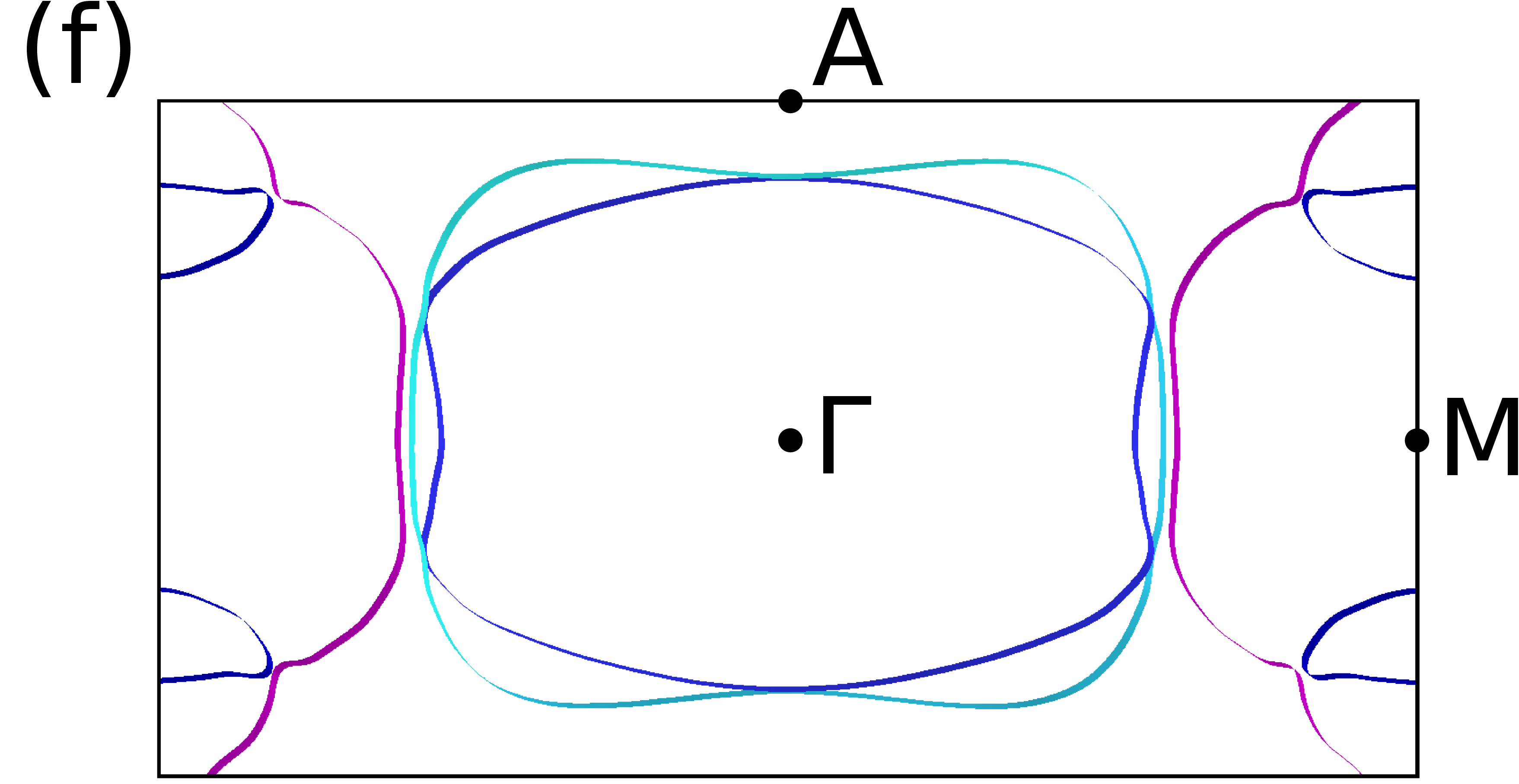}
\caption{
\label{fig:fermi_surface}
Fermi surface sheets of $\varepsilon$-Fe obtained by DFT+DMFT at 12~GPa [(a)-(d)]. Plots (e) and (f) show cuts of the Fermi surface sheets by the planes
passing through points M,$\Gamma$,K and  M,$\Gamma$,A, respectively. The line width has no meaning.}
\end{figure*}

To study possible symmetries of electron pairing, we consider simplified approach, using pairing interaction vertex $V_{{\bf k}{\bf k^{\prime}},\alpha\beta}^{s,t}$ ($\alpha,\beta$ are band indices, ${\bf k}$,${\bf k}'$ are the respective momenta) in the singlet (s) and triplet (t) channels, obtained in the second order of perturbation theory, similarly to the original Kohn and Luttinger paper \cite{KohnSC}, see also Refs. \cite{KohnSC1,Sr2RuO4} and Appendix. Although the Coulomb interaction in $\varepsilon$-iron is not small, we expect that this approach correctly reproduces the symmetry of the gap function, in view of the Fermi-liquid behavior of electronic self-energies at low temperatures, obtained in Sect. \ref{Sect:elProp}. For more sophisticated approaches, accounting for higher-order diagrams, e.g., in a ladder series \cite{Scalapino1987,SE_DF,HeldSC}, in the presence of strong electronic correlations, it is preferable
considering dynamic vertices \cite{OurRevModPhys} instead of static ones, which requires introducing Moriyaesque lambda correction \cite{KataninMoriya} and/or treating non-local self-energy feedback \cite{SE_DF}
to avoid unphysical divergences of the series.
Such generalizations previously were explored for the single-band models only (see, e.g., Refs. \cite{SE_DF,HeldSC}). For the considering multi-band models they are rather tedious and not performed here.

To determine pairing symmetry we consider eigenvalues $\lambda^{s,t}$ and eigenfunctions $f^{s,t}_{{\bf k},\alpha}$ of the Bethe-Salpeter equation 
\begin{equation}
\lambda^{s,t}f^{s,t}_{{\bf k},\alpha}   =-T\!\!\sum_{{\bf k}^{\prime},i\nu_n,\beta}\!V_{{\bf k}{\bf k}^{\prime},\alpha\beta}^{s,t}
f_{{\bf k}^{\prime}\beta}^{s,t} G_{{\bf k'}\beta} (i\nu_n) G_{-{\bf k'}\beta}(-i\nu_n)%
\label{EqGap3}
\end{equation}
where 
$G_{{\bf k}\alpha}(i\nu_n)$ is the Green function for the band $\alpha$, obtained by the diagonalization of the Green functions matrices $G_{\bf k}(i\nu_n)$ with respect to orbital indexes.
We parametrize the gap functions $f_{{\bf k},\alpha}$ for momenta lying on the Fermi surface and assume them to be independent on the radial direction of ${\bf k}$, since the summation in Eq.~(\ref{EqGap3}) is anyhow dominated by the vicinity of the Fermi surface. Among many sheets of the Fermi surface (see Fig.~\ref{fig:fermi_surface}) we choose the two most important [Figs.~\ref{fig:fermi_surface}(a) and \ref{fig:fermi_surface}(b)], which parts are connected by nesting vectors. We parametrize points of these Fermi surface sheets by the polar and azimuthal angles $(\theta,\phi)$, where ${\phi=0}$ corresponds to point M of the Brillouin zone, ${\theta=0,\pi}$ to the ``North" and ``South" poles. 

\begin{figure}[b]
\includegraphics[clip=true,width=0.415\textwidth]{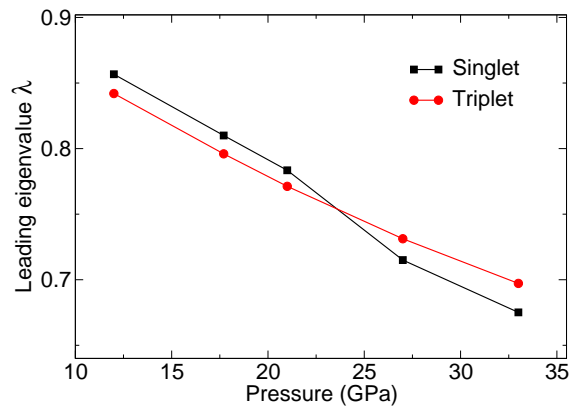}
\caption{Leading eigenvalue of the Bethe-Salpeter equation for singlet and triplet pairing in $\varepsilon$-Fe calculated at ${\beta=100}$~eV$^{-1}$.
\label{fig:eigenval}}
\end{figure}

\begin{figure*}[t]
\includegraphics[clip=true,width=0.4\textwidth]{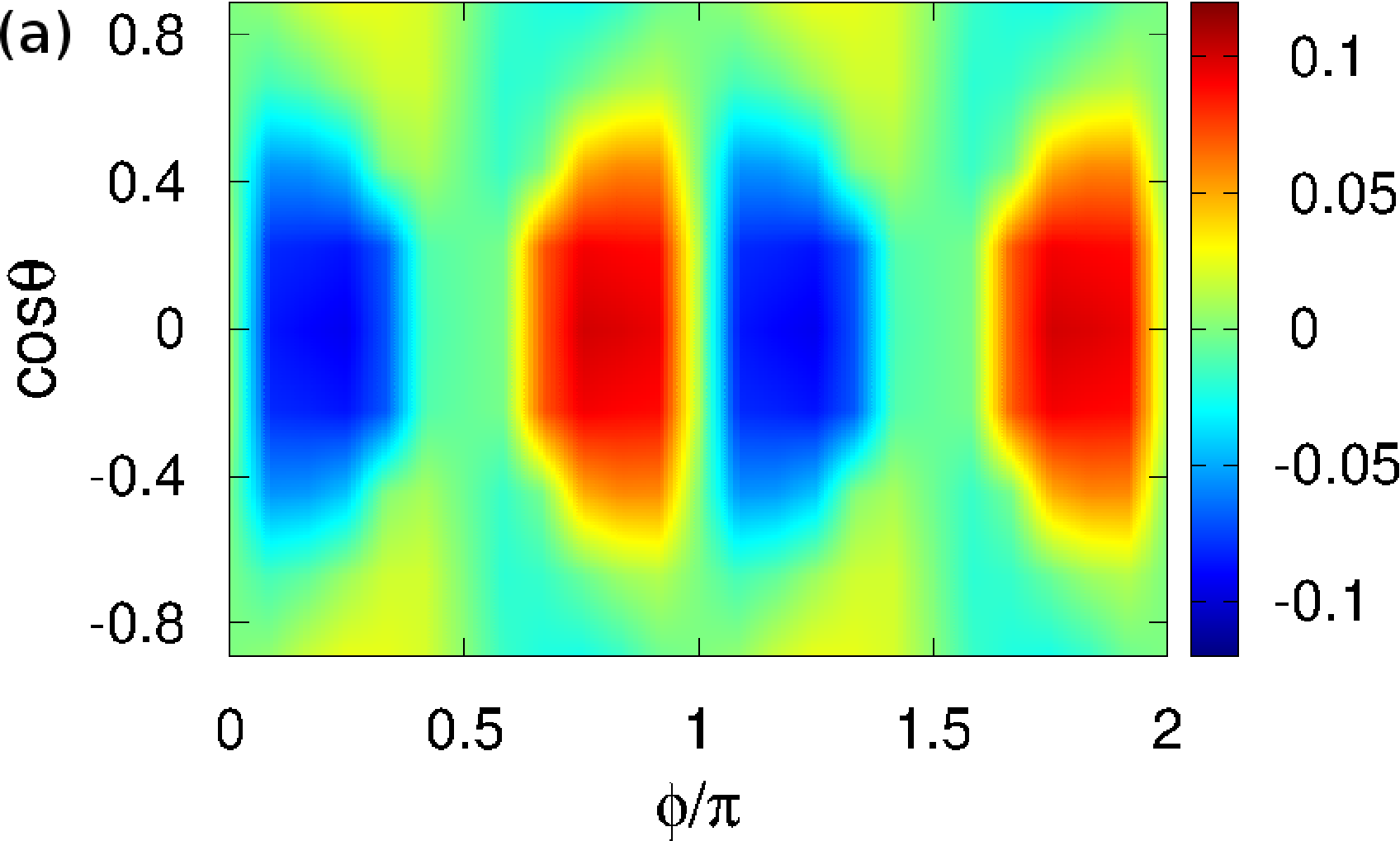}
\hspace{0.6cm}
\includegraphics[clip=true,width=0.4\textwidth]{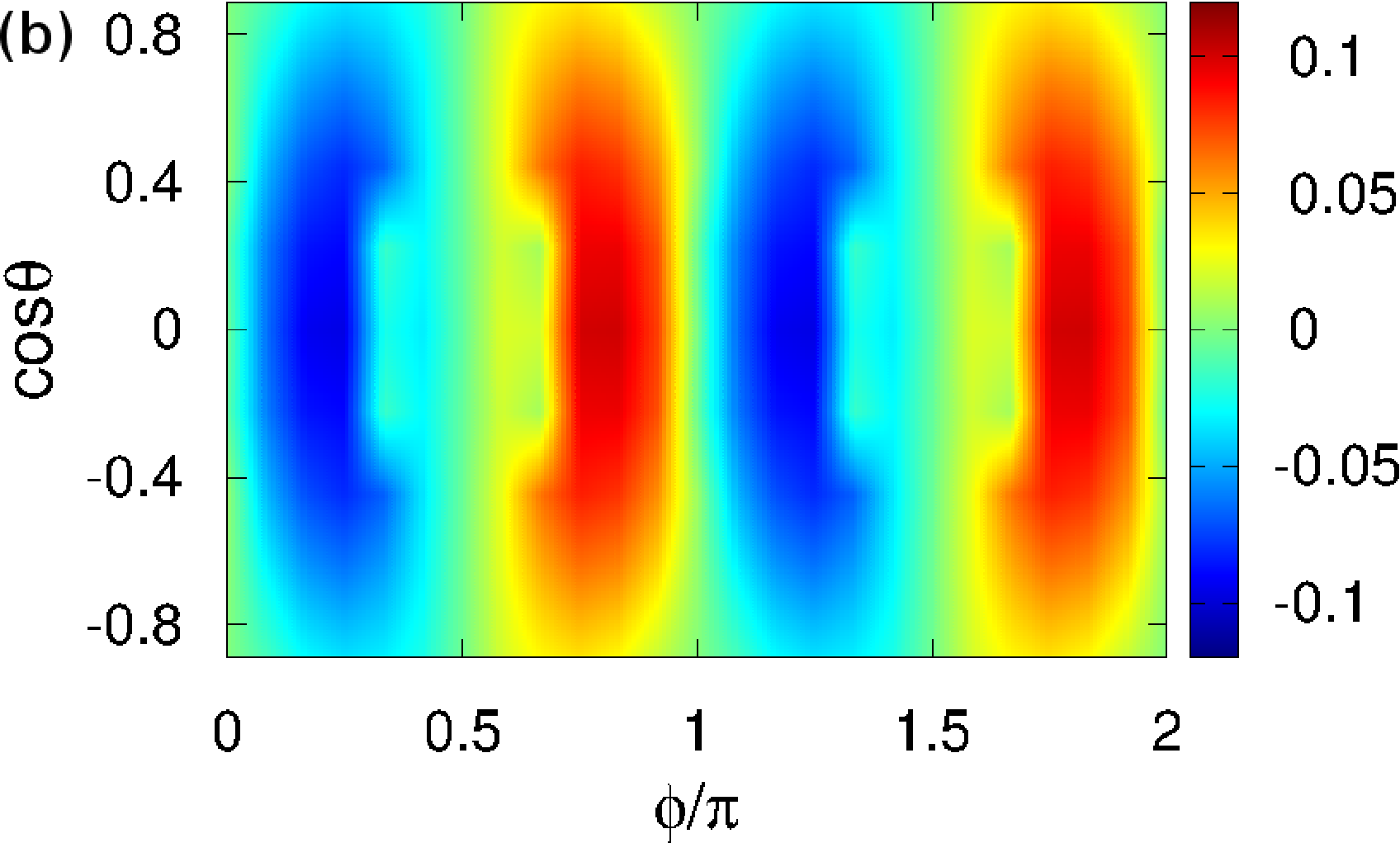}\\
\vspace{0.15cm}
\includegraphics[clip=true,width=0.4\textwidth]{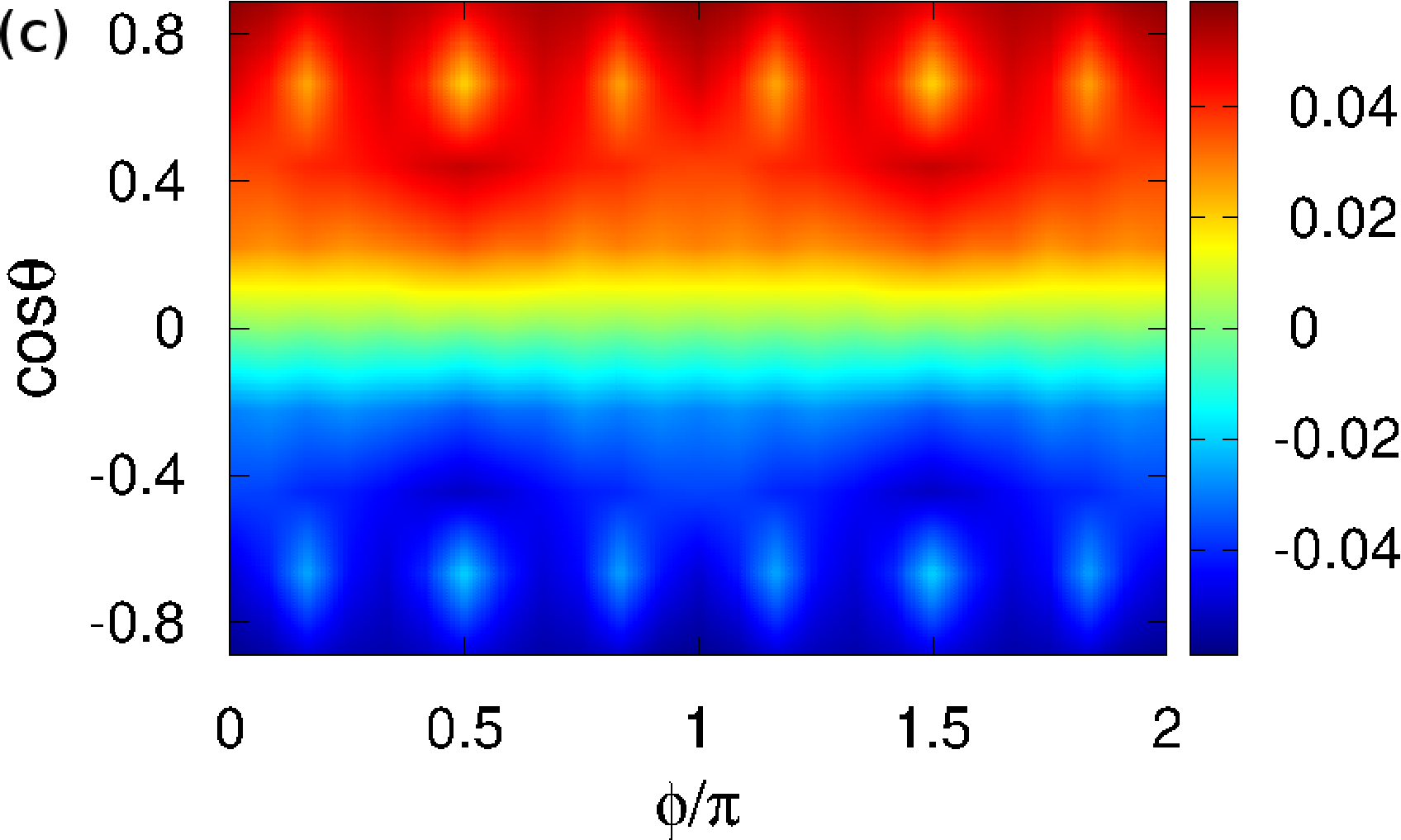}
\hspace{0.6cm}
\includegraphics[clip=true,width=0.4\textwidth]{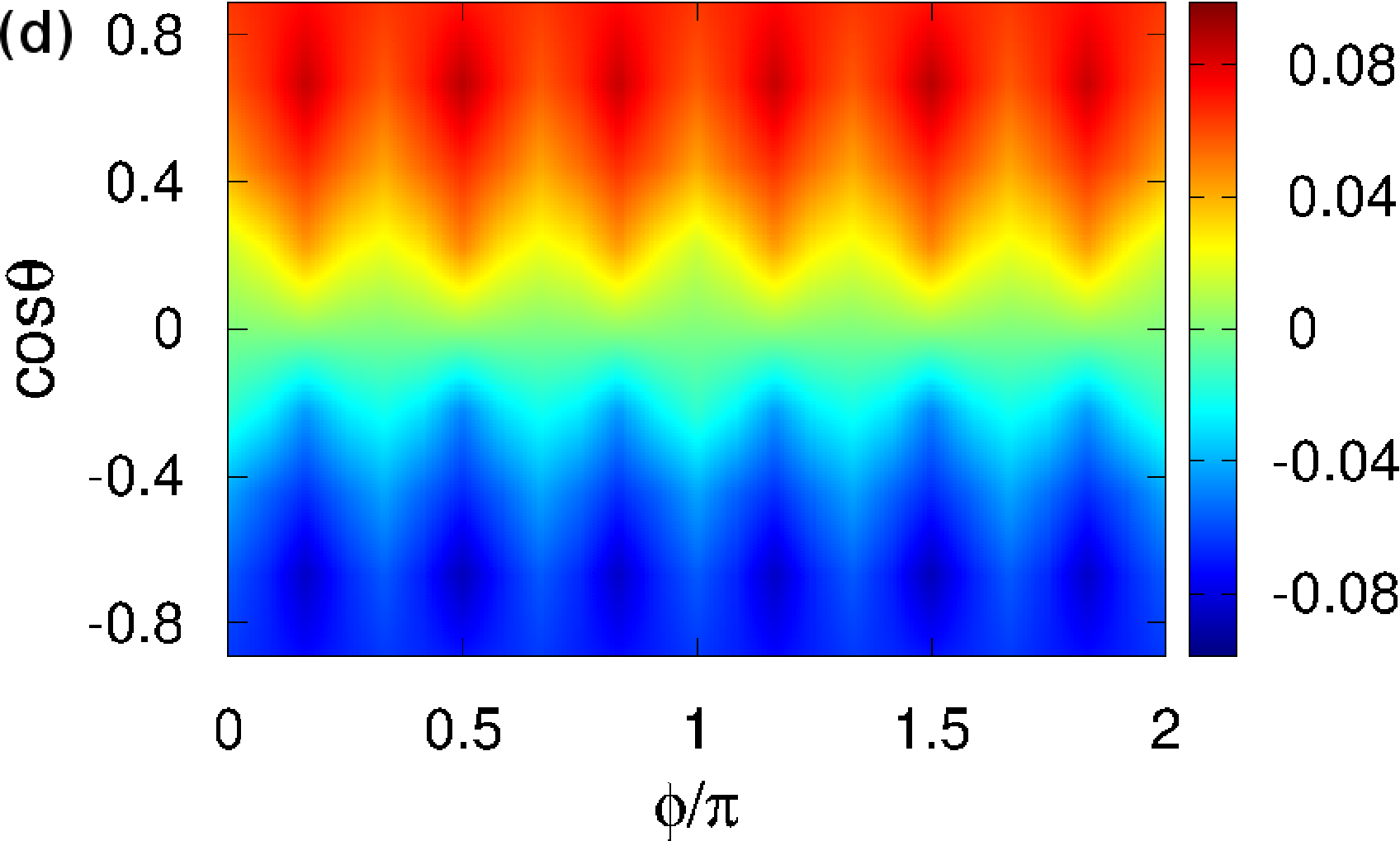}
\caption{
Eigenfunction of the Bethe-Salpeter
equation for singlet (a,b) and triplet (c,d) pairing as a function of polar $\theta$ and azimuthal $\phi$ angles
calculated at pressure of 21 GPa and $\beta=100$~eV$^{-1}$.
(a), (c) and (b), (d) correspond to sheets of Fermi surface 
depicted in Figs.~\ref{fig:fermi_surface}(a) and \ref{fig:fermi_surface}(b), respectively. 
Angle ${\phi=0}$ corresponds to point M of the Brillouin zone.
\label{fig:gap}}
\end{figure*}

The largest (leading) eigenvalue determines the preferred momentum dependence of the pairing gap, given by the corresponding eigenfunction \cite{Bulut1993,NotePairing}. Calculated leading eigenvalues of singlet and triplet pairing are shown in Fig.~\ref{fig:eigenval}. For the considered range of pressures, singlet and triplet eigenvalues are close to each other. Interestingly, we obtain crossing of eigenvalues of singlet and triplet pairing near the pressure $p_c=22$~GPa, close to that, at which maximum value of superconducting $T_c$ is obtained. Yet, both eigenvalues increase with decreasing pressure, so that the obtained pressure dependence agrees with the experimental data for $p>p_c$. For $p<p_c$ the considered theory does not describe the suppression of superconductivity with decreasing pressure. The increase of the eigenvalues with decreasing pressure, however, goes along with the increase of the non-uniform susceptibility, see Fig.~\ref{fig:susc_irr} and shows the spin-fluctuation origin of the obtained superconducting order parameters. We note that the obtained close competition of singlet and triplet pairing is similar to that discussed previously for Sr$_2$RuO$_4$ (see, e.g., Refs.~\cite{Sr2RuO4,Sr2RuO4_1}) and may be lifted by going beyond second-order perturbation theory, including nearest-neighbor Coulomb repulsion, spin-orbit coupling etc.

Obtained momentum dependence of the gap functions for singlet order parameters [Figs.~\ref{fig:gap}(a) and \ref{fig:gap}(b)] take the values $\approx0,\pm \Delta$ at the subsequent vertical faces of the considered Fermi surface sheets. This momentum dependence allows one to achieve a change of the sign of gap function on the opposite sides of the Fermi surface, connected by the wave vector of preferred incommensurate fluctuations in the K-M direction. On the contrary, the triplet pairing gap is almost $\phi$-independent and changes sign at ${\theta=\pi/2}$, having a single line of the nodes.

\section{Conclusions}
\label{sec:conclusions}

We have studied electronic and magnetic properties of $\varepsilon$-Fe at pressures up to 33~GPa by employing the DFT+DMFT approach and using its results as a starting point for the multi-band Bethe-Salpeter equation. Our main results 
are as follows.

(i) We have not found a significant deviation from the FL behavior below 1000~K even with relatively large ${U=6}$~eV. This points to the importance of non-local correlations effects, neglected in DMFT.

(ii) At the same time, the Coulomb correlations lead to substantial electron mass enhancement up to $m^*/m\sim 1.8$, which is accompanied by renormalization of the density of states. As a result, the distance from the peak lying above the Fermi level is decreased to 0.4~eV compared to 0.75~eV obtained in DFT calculations.

(iii) We have found short-lived magnetic moments with the lifetime $\sim10$ fs, comparable to that estimated previously in pnictides. This may explain the discrepancy of XES experiments with the M\"ossbauer spectroscopy because of the longer time scales of the latter. Namely, the lifetime of local magnetic moments or spin fluctuations can be small enough to be undetected by the M\"ossbauer measurements.

(iv) We find that the antiferromagnetic spin fluctuations are the dominant ones. At the same time, we do not find substantial ferromagnetic correlations responsible for the $T^{5/3}$ dependence of the resistivity. This indicates that the non-local correlation effects beyond DMFT may play a significant role in $\varepsilon$-Fe. In particular, non-local correlation effects corresponding to virtual transitions into a peak lying near the Fermi level may give a contribution \cite{Katsnelson}. 
For this contribution the peak needs not to lie directly at $E_F$; moreover, its distance from $E_F$ needs not to be too small.

(v) The antiferromagnetic spin fluctuations in $\varepsilon$-Fe can 
result
in the superconducting instability.
%
The obtained results allow us to explain the appearance of superconducting instability in the considered temperature range and its disappearance with pressure, but not 
a gradual grow of the superconducting critical temperature $T_c$ from 15 to 21~GPa.
%
%
The growth of $T_c$ with pressure in the low-pressure range may be 
explained by earlier suggested ~\cite{Jarlborg} presence of magnetic clusters of $\alpha$-phase within \e-phase, which both are metastable 
in a wide pressure range.
These magnetic clusters, more common at low pressures,
or, alternatively, other lattice defects can affect the electron pairing, leading to suppression of the critical temperature.
%
The latter explanation is supported by the experimental fact 
that $\alpha\!\rightarrow\!\varepsilon$ martensitic transition
is sluggish,
and pressures higher than 20~GPa are needed for a complete transformation~\cite{Basset1987,Jaccard2002}.

As it follows from the above discussion, further study of the non-local correlations in $\varepsilon$-iron seems to be important. This would allow studying in more detail mechanism of superconductivity and its pressure dependence, and may allow to explain observed $T^{5/3}$ behavior of resistivity, accompanied by a quantum phase transition at the pressure $p=21$~GPa \cite{Yadav2013}. On the other hand, further experimental studies are required to clarify the magnetic and superconducting properties of \e-Fe, that may also have significant implications for geophysics and Earth's core magnetism.





\vspace{-0.1cm}
\begin{acknowledgments}
\vspace{-0.01cm}
The DFT+DMFT calculations and solution of the BSE were supported by Russian Science Foundation (project 19-12-00012).
The calculations of the particle-hole bubble were supported by the Ministry of Science and Higher Education of the Russian Federation (theme “Electron” No. AAAA-A18-118020190098-5).
\end{acknowledgments}

\onecolumngrid

\appendix
\section{Superconducting pairing}
\subsection{General formalism and the gap equation}

We consider the general form of the non-interacting Hamiltonian
\begin{equation}
\hat{H}_{0}=\sum\limits_{{\bf k}\sigma}\sum\limits_{mm^{\prime}}\varepsilon
_{\bf k}^{mm^{\prime}}c_{{\bf k},m\sigma}^{\dagger}c_{{\bf k},m\sigma}%
\end{equation}
(we include the real part of the self-energies at zero frequency and double counting into the definition of $\varepsilon_{\bf k}^{mm^{\prime}}$) and $4$-fermion interaction
\begin{align}
\hat{H}_{\mathrm{int}}  & =-\frac{1}{4}\sum\limits_{\sigma\sigma^{\prime
}\sigma^{\prime\prime}\sigma^{\prime\prime\prime}}\sum\limits_{mm^{\prime
}m^{\prime\prime}m^{\prime\prime\prime}}\sum\limits_{{\bf k}_{1}{\bf k}_{2}{\bf k}_{3}{\bf k}_{4}%
}\Gamma_{{\bf k}_{1},{\bf k}_{2};{\bf k}_{3},{\bf k}_{4}}^{\sigma\sigma^{\prime}\sigma^{\prime\prime
}\sigma^{\prime\prime\prime},mm^{\prime}m^{\prime\prime}m^{\prime\prime\prime
}}c_{{\bf k}_{1},m\sigma}^{\dagger}c_{{\bf k}_{2},m^{\prime}\sigma^{\prime}}^{\dagger
}c_{{\bf k}_{3},m^{\prime\prime}\sigma^{\prime\prime}}c_{{\bf k}_{4},m^{\prime\prime
\prime}\sigma^{\prime\prime\prime}}\nonumber\\
& \times\delta({\bf k}_{1}+{\bf k}_{2}-{\bf k}_{3}-{\bf k}_{4}),
\label{H4}
\end{align}
where $m,m^{\prime},m^{\prime\prime},m^{\prime\prime\prime}$ are the orbital
indexes, $\sigma,\sigma^{\prime},\sigma^{\prime\prime},\sigma^{\prime
\prime\prime}=\uparrow,\downarrow$ are the spin indexes, $c_{{\bf k}m\sigma}$ are the fermionic operators.
Due to the Pauli principle,
the interaction vertex $\Gamma$ is antisymmetric:
\begin{subequations}
\begin{align}
&  \Gamma_{{\bf k}_{1},{\bf k}_{2};{\bf k}_{3},{\bf k}_{4}}^{\sigma\sigma^{\prime}\sigma^{\prime
\prime}\sigma^{\prime\prime\prime},mm^{\prime}m^{\prime\prime}m^{\prime
\prime\prime}}=V_{{\bf k}_{1},{\bf k}_{2};{\bf k}_{3},{\bf k}_{4}}^{mm^{\prime}m^{\prime\prime
}m^{\prime\prime\prime}}\delta_{\sigma\sigma^{\prime\prime}}\delta
_{\sigma^{\prime}\sigma^{\prime\prime\prime}}-V_{{\bf k}_{1},{\bf k}_{2};{\bf k}_{4},{\bf k}_{3}%
}^{mm^{\prime}m^{\prime\prime\prime}m^{\prime\prime}}\delta_{\sigma
\sigma^{\prime\prime\prime}}\delta_{\sigma^{\prime}\sigma^{\prime\prime}}\label{VV}\\
&  V_{{\bf k}_{1},{\bf k}_{2};{\bf k}_{3},{\bf k}_{4}}^{mm^{\prime}m^{\prime\prime}m^{\prime
\prime\prime}}=V_{{\bf k}_{2},{\bf k}_{1};{\bf k}_{4},{\bf k}_{3}}^{m^{\prime}m,m^{\prime\prime\prime
}m^{\prime\prime}}=\left(  V_{{\bf k}_{3},{\bf k}_{4};{\bf k}_{1},{\bf k}_{2}}^{m^{\prime\prime
}m^{\prime\prime\prime},mm^{\prime}}\right)  ^{\ast},
\end{align}
where for $SU(2)$ symmetric interaction the vertices $V$ do not depend on spin indexes.

To discuss superconducting pairing, we perform the mean-field decoupling 
\end{subequations}
\begin{equation}
\hat{H}_{\mathrm{int}}\rightarrow\hat{H}_{\mathrm{int}}^{\mathrm{BCS}}%
=-\frac{1}{4}\sum\limits_{{\bf k}{\bf p},mm^{\prime},m^{\prime\prime}m^{\prime\prime
\prime}}\sum\limits_{\sigma\sigma^{\prime}\sigma^{\prime\prime}\sigma
^{\prime\prime\prime}}\left[  \Gamma_{{\bf k},-{\bf k};{\bf p},-{\bf p}}^{\sigma\sigma^{\prime}%
\sigma^{\prime\prime}\sigma^{\prime\prime\prime},mm^{\prime}m^{\prime\prime
}m^{\prime\prime\prime}}c_{{\bf k},m\sigma}^{\dagger}c_{-{\bf k},m'\sigma^{\prime}%
}^{\dagger}\langle c_{{\bf p},m^{\prime\prime}\sigma^{\prime\prime}}c_{-{\bf p},m^{\prime
\prime\prime}\sigma^{\prime\prime\prime}}{\rangle}+h.c.\right]  .
\end{equation}
We suppose that the quadratic part $\hat{H}_{0}$ of the Hamiltonian is
diagonalized by the transformation $c_{{\bf p},m\sigma}=\sum\limits_{\alpha}%
U_{{\bf p}}^{m\alpha}e_{{\bf p},\alpha\sigma}$ where $e_{{\bf k},\alpha\sigma}$ are the fermionic operators for the band $\alpha$ and spin projection $\sigma$, such
that
\begin{equation}
\hat{H}_{0}=\sum\limits_{{\bf k}\alpha\sigma}\epsilon_{{\bf k}\alpha}e_{{\bf k},\alpha\sigma
}^{\dagger}e_{{\bf k},\alpha\sigma},
\end{equation}
where $E_{{\bf k}\alpha}=\sum\nolimits_{mm^{\prime}}\varepsilon_{\bf  k}^{mm^{\prime}%
}(U_{\bf  k}^{m\alpha})^{\ast}U_{\bf k}^{m^{\prime}\alpha},$ $\sum\nolimits_{mm^{\prime
}}\varepsilon_{\bf k}^{mm^{\prime}}(U_{\bf k}^{m\alpha})^{\ast}U_{\bf k}^{m^{\prime}\beta
}=0$ ($\alpha\neq\beta$). Then we obtain
\begin{align}
\hat{H}_{\mathrm{int}}^{\mathrm{BCS}} &  =-\frac{1}{4}\sum\limits_{\bf kp}%
\sum\limits_{\sigma\sigma^{\prime}\sigma^{\prime\prime}\sigma^{\prime
\prime\prime},\alpha\beta}\left[  \Gamma_{{\bf k},{\bf p}}^{\sigma\sigma^{\prime}%
\sigma^{\prime\prime}\sigma^{\prime\prime\prime},\alpha\beta}e_{{\bf k},\alpha
\sigma}^{\dagger}e_{-{\bf k},\alpha\sigma^{\prime}}^{\dagger}\langle e_{{\bf p},\beta
\sigma^{\prime\prime}}e_{-{\bf p},\beta\sigma^{\prime\prime\prime}}{\rangle
}+h.c.\right]  \nonumber\\
&  =\frac{1}{2}\sum\limits_{{\bf k}\alpha,\sigma\sigma^{\prime}}\left[
\Delta_{{\bf k},\alpha}^{\sigma\sigma^{\prime}}e_{{\bf k},\alpha\sigma}^{\dagger
}e_{-{\bf k},\alpha\sigma^{\prime}}^{\dagger}+h.c.\right]  ,
\end{align}
where
\begin{equation}
\Delta_{{\bf k},\alpha}^{\sigma,\sigma^{\prime}}=-\frac{1}{2}\sum\limits_{{\bf p},\sigma
^{\prime\prime}\sigma^{\prime\prime\prime},\beta}\Gamma_{{\bf k},{\bf p}}^{\sigma
\sigma^{\prime}\sigma^{\prime\prime}\sigma^{\prime\prime\prime},\alpha\beta
}\langle e_{{\bf p},\beta\sigma^{\prime\prime}}e_{-{\bf p},\beta\sigma^{\prime\prime
\prime}}{\rangle}%
\end{equation}
and%
\begin{equation}
\Gamma_{{\bf k},{\bf p}}^{\sigma\sigma^{\prime}\sigma^{\prime\prime}\sigma^{\prime
\prime\prime},\alpha\beta}=\sum\limits_{mm^{\prime},m^{\prime\prime}%
m^{\prime\prime\prime}}\Gamma_{{\bf k},-{\bf k};{\bf p},-{\bf p}}^{\sigma\sigma^{\prime}\sigma
^{\prime\prime}\sigma^{\prime\prime\prime},mm^{\prime}m^{\prime\prime
}m^{\prime\prime\prime}}\left(  U_{\bf {\bf k}}^{m\alpha}\right)  ^{\ast}%
(U_{-{\bf k}}^{m^{\prime}\alpha})^* U_{{\bf p}}^{m^{\prime\prime}\beta} U_{-{\bf p}}%
^{m^{\prime\prime\prime}\beta}.
\end{equation}
As usually, we separate singlet and triplet terms:
\begin{align}
\Delta_{{\bf k},\alpha}^{\sigma,\sigma^{\prime}} &  =-\Delta_{-{\bf k},\alpha}%
^{\sigma^{\prime},\sigma}=(\Delta_{{\bf k},\alpha}^{\sigma,\sigma^{\prime}}%
)_{s}+(\Delta_{{\bf k},\alpha}^{\sigma,\sigma^{\prime}})_{t},\\
(\Delta_{{\bf k},\alpha}^{\sigma,\sigma^{\prime}})_{s} &  =(\Delta_{-{\bf k},\alpha
}^{\sigma,\sigma^{\prime}})_{s}=-(\Delta_{{\bf k},\alpha}^{\sigma^{\prime},\sigma
})_{s}=i\sigma_{y}^{\sigma\sigma^{\prime}}\psi_{{\bf k},\mu}=\left(
\begin{array}
[c]{cc}%
0 & \psi_{{\bf k},\alpha}\\
-\psi_{{\bf k},\alpha} & 0
\end{array}
\right)  _{\sigma\sigma^{\prime}};\nonumber\\
(\Delta_{{\bf k},\alpha}^{\sigma,\sigma^{\prime}})_{t} &  =-(\Delta_{-{\bf k},\alpha
}^{\sigma,\sigma^{\prime}})_{t}=(\Delta_{{\bf k},\alpha}^{\sigma^{\prime},\sigma
})_{t}=i({\bf{d}}_{{\bf k},\mu}{\mbox {\boldmath $\sigma $}}^{\sigma\sigma^{\prime\prime}})\sigma
_{y}^{\sigma^{\prime\prime}\sigma^{\prime}}=\left(
\begin{array}
[c]{cc}%
-d_{{\bf k},\alpha}^{x}+id_{{\bf k},\alpha}^{y} & d_{k,\alpha}^{z}\\
d_{{\bf k},\alpha}^{z} & d_{{\bf k},\alpha}^{x}+id_{{\bf k},\alpha}^{y}%
\end{array}
\right)  _{\sigma\sigma^{\prime}}.\nonumber
\end{align}
where $\mbox {\boldmath $\sigma $}$ are the Pauli matrices. From these relations, we find%
\begin{align}
(\Delta_{{\bf k},\alpha}^{\sigma,\sigma^{\prime}})_{s(t)} &  =\frac{1}{2}\left(
\Delta_{{\bf k},\alpha}^{\sigma,\sigma^{\prime}}\mp\Delta_{{\bf k},\alpha}^{\sigma
^{\prime},\sigma}\right)  =-\frac{1}{2}\sum\limits_{{\bf p},\sigma^{\prime\prime
}\sigma^{\prime\prime\prime},\beta}\Gamma_{s(t),{\bf k},{\bf p}}^{\sigma\sigma^{\prime
}\sigma^{\prime\prime}\sigma^{\prime\prime\prime},\alpha\beta}\langle
e_{{\bf p},\beta\sigma^{\prime\prime}}e_{-{\bf p},\beta\sigma^{\prime\prime\prime}%
}{\rangle}\nonumber\\
&  =-\sum\limits_{{\bf p},\beta
}V_{{\bf k}{\bf p},\alpha\beta}^{s(t)}\langle e_{{\bf p},\beta\sigma}%
e_{-{\bf p},\beta\sigma^{\prime}}{\rangle,}%
\label{MF_eq}
\end{align}
where
\begin{align}
\Gamma_{s(t),{\bf k},{\bf p}}^{\sigma\sigma^{\prime}\sigma^{\prime\prime}\sigma
^{\prime\prime\prime},\alpha\beta}  & =\frac{1}{2}\left(  \Gamma_{{\bf k},{\bf p}}%
^{\sigma\sigma^{\prime}\sigma^{\prime\prime}\sigma^{\prime\prime\prime}%
,\alpha\beta}\mp\Gamma_{{\bf k},{\bf p}}^{\sigma^{\prime}\sigma\sigma^{\prime\prime}%
\sigma^{\prime\prime\prime},\alpha\beta}\right)  ,\\
V_{{\bf k,p},\alpha\beta}^{s(t)}  & =\frac{1}{2}\sum\limits_{mm^{\prime},m^{\prime\prime
}m^{\prime\prime\prime}}\left(  V_{{\bf k},-{\bf k};{\bf p},-{\bf p}}^{mm^{\prime}m^{\prime\prime
}m^{\prime\prime\prime}}\pm V_{{\bf k},-{\bf k};-{\bf p},{\bf p}}^{mm^{\prime}m^{\prime\prime\prime
}m^{\prime\prime}}\right) \left(  U_{\bf k}^{m\alpha}\right)  ^{\ast}%
(U_{-{\bf k}}^{m^{\prime}\alpha})^* U_{{\bf p}}^{m^{\prime\prime}\beta} U_{-{\bf p}}%
^{m^{\prime\prime\prime}\beta} .\label{Vst}
\end{align}
From Eq. (\ref{MF_eq}) we obtain gap equations:
\begin{subequations}
\label{EqGap}%
\begin{align}
\psi_{{\bf k},\alpha} &  =-T\sum_{{\bf k}^{\prime},i\nu'_n,\beta}V_{{\bf k}{\bf k}^{\prime},\alpha\beta}^{s}%
F^{s}_{{\bf k}'\beta}(i\nu'_n)\label{EqGap1}\\
{\bf d}_{{\bf k},\alpha} &  =-T\sum_{{\bf k}^{\prime},i\nu'_n,\beta}V_{{\bf k},{\bf k}^{\prime},\alpha\beta}%
^{t}{\bf F}^{t}_{{\bf k}'\beta}(i\nu'_n),\label{EqGap2}
\end{align}
where $F^{s}_{{\bf k}\alpha}(i\nu_n)=(i\sigma_y^{\sigma \sigma'}/2)\langle
e_{{\bf k}\alpha\sigma}(\tau) e_{{\bf k}\alpha\sigma'}(0)
\rangle_{i \nu_n} $ and ${\bf F}^{t}_{{\bf k}\alpha}(i\nu_n)=(i/2)(\mbox {\boldmath $\sigma $}\sigma
_{y})^{\sigma\sigma^{\prime}}\langle
e_{{\bf k}\alpha\sigma}(\tau) e_{{\bf k}\alpha\sigma'}(0)
\rangle_{i \nu_n} $ are the anomalous Green functions of the Hamiltonian ${\hat H}_0+{\hat H}^{\rm BCS}_{\rm int}$, the index ${i \nu_n}$ denotes the Fourier transform.
The corresponding Bethe-Salpeter equations are obtained by linearizing the gap equations and considering the eigenvalues and eigenfunctions of the operators in the right-hand-side of Eq. (\ref{EqGap}), see Eq. (\ref{EqGap3}) of the main text.
\end{subequations}

\subsection{Bare and second-order vertex}

To determine the vertex $V,$ we use the second-order perturbation theory.
%
To preserve spin-rotational symmetry, we consider the following orbital-dependent interactions%
\begin{align}
H_{\text{int}}&=U\sum\limits_{im}n_{im\uparrow}n_{im\downarrow}+\sum
\limits_{i,m>m^{\prime},\sigma}U_{1}^{mm^{\prime}}n_{im\sigma}n_{im^{\prime
}\sigma}+\sum\limits_{i,m>m^{\prime},\sigma}U_{2}^{mm^{\prime}}n_{im\sigma
}n_{im^{\prime},-\sigma}\notag\\%
& +\sum\limits_{i,m>m^{\prime},\sigma\sigma^{\prime}}(U_{2}^{mm^{\prime}%
}-U_{1}^{mm^{\prime}})c_{im\sigma}^{+}c_{im,-\sigma}c_{im^{\prime},-\sigma
}^{+}c_{im^{\prime},\sigma}+\sum\limits_{i,m\ne m^{\prime}}J^{mm^{\prime}%
}c_{im\uparrow}^{+}c^{+}_{im,\downarrow}c_{im^{\prime},\downarrow
}c_{im^{\prime},\uparrow}%
\label{H_multiorb}
\end{align}
(for the standard Kanamori parametrization with Hund exchange $I$, interorbital exchange $U'-I/2$, and pair hopping $J$ we have $U_{2}^{mm^{\prime}}=U'$, $U_{1}^{mm^{\prime}}=U'-I$, $J^{mm^{\prime}}=J$).
For the bare local vertex $V_{0}^{mm^{\prime}m^{\prime\prime}%
m^{\prime\prime\prime}}$ defined by putting the Hamiltonian (\ref{H_multiorb}) into the form of 
Eqs. (\ref{H4}) and (\ref{VV}) we have%
\begin{align}
V_{0}^{mm^{\prime},m^{\prime\prime}m^{\prime\prime\prime}}  & =U\delta
_{mm^{\prime\prime}}\delta_{m^{\prime}m^{\prime\prime\prime}}\delta
_{mm^{\prime}}+U_{2}^{mm^{\prime}}\delta_{mm^{\prime\prime}}\delta_{m^{\prime
}m^{\prime\prime\prime}}\left(  1-\delta_{mm^{\prime}}\right)  \notag \\
& +(U_{2}^{mm^{\prime}}-U_{1}^{mm^{\prime}})\delta_{mm^{\prime\prime\prime}%
}\delta_{m^{\prime}m^{\prime\prime}}\left(  1-\delta_{mm^{\prime}}\right)+J^{mm^{\prime\prime}}\delta_{mm^{\prime}%
}\delta_{m^{\prime\prime}m^{\prime\prime\prime}}\left(  1-\delta_{mm^{\prime\prime}}\right)  .
\end{align}
In the second order in $V_{0}$ the effective interaction is given by $V_{0}^{mm^{\prime}m^{\prime\prime}m^{\prime\prime\prime}}+V_{2,{\bf k},-{\bf k};{\bf p},-{\bf p}}^{mm^{\prime}m^{\prime\prime}m^{\prime\prime\prime}}$ where%
\begin{align}
V_{2,{\bf k},-{\bf k};{\bf p},-{\bf p}}^{mm^{\prime}m^{\prime\prime}m^{\prime\prime\prime}} &
=L_{ph}^{\widetilde{m}^{\prime\prime}\widetilde{m}^{\prime\prime\prime
},\widetilde{m}^{\prime}\widetilde{m}}({\bf k}-{\bf p})\left[  -2V_{0}^{m\widetilde
{m},m^{\prime\prime}\widetilde{m}^{\prime\prime}}V_{0}^{\widetilde{m}^{\prime
}m^{\prime},\widetilde{m}^{\prime\prime\prime}m^{\prime\prime\prime}}%
+V_{0}^{m\widetilde{m},\widetilde{m}^{\prime\prime}m^{\prime\prime}}%
V_{0}^{\widetilde{m}^{\prime}m^{\prime},\widetilde{m}^{\prime\prime\prime
}m^{\prime\prime\prime}}\right.  \nonumber\\
&  \left.  +V_{0}^{m\widetilde{m},m^{\prime\prime}\widetilde{m}^{\prime\prime
}}V_{0}^{\widetilde{m}^{\prime}m^{\prime},m^{\prime\prime\prime}\widetilde
{m}^{\prime\prime\prime}}\right]+L_{ph}^{\widetilde{m}\widetilde{m}^{\prime
},\widetilde{m}^{\prime\prime}\widetilde{m}^{\prime\prime\prime}}%
({\bf k}+{\bf p})V_{0}^{m\widetilde{m}^{\prime\prime\prime},\widetilde{m}m^{\prime
\prime\prime}}V_{0}^{\widetilde{m}^{\prime\prime}m^{\prime},m^{\prime\prime
}\widetilde{m}^{\prime}}  ,\label{A22}
\end{align}
where we assume summation over repeating orbital indices and%
\begin{equation}
L_{ph}^{mm^{\prime},m^{\prime\prime}m^{\prime\prime\prime}}({\bf q})=-T\sum
\limits_{{\bf k},i\nu_n}G_{\bf k}^{mm^{\prime\prime}}(i\nu_n)G_{{\bf k}+{\bf q}}^{m^{\prime}m^{\prime\prime\prime}}(i\nu_n)%
\end{equation}
is the particle-hole bubble. 




\twocolumngrid


\begin{thebibliography}{9}
\bibitem{NotePairing} The eigenvalue $\lambda$=1 corresponds to the pairing instability. In view of our qualitative analysis of the second-order perturbation theory we find only the eigenfunctions, corresponding to the leading eigenvalue, and do not obtain corresponding pairing temperatures.
\end{thebibliography}
\end{document}